# A Study of Network Based Mobility Management Schemes, 6LoWPAN Mobility, Open Issues and Proposed Solutions.

Riaz A Khan[a], A H Mir[b]

[a,b]Department of Electronics and Communication Engineering,

[a,b]National Institute of Technology (NIT), Srinagar, Kashmir-190006, India,

Corresponding Author; Email address: riazk3@gmail.com, Tel.: +91-9469433336.

**Abstract:** Wireless Sensor Nodes (SNs), the key elements for building Internet of Things (IOT), have been deployed widely in order to get and transmit information over the internet. With the introduction of IPv6 over Low Power Wireless Personal Area Network (6LoWPAN), it is possible to connect these constrained devices to IPv6 Networks and transmit IPv6 packets. The sensor nodes are being deployed/installed on many objects and some of them are mobile (moving) including mobile gadgets, physical objects (living or non-living) etc. These mobile objects require sufficient Mobility Management Schemes to take care of data transmission. Host based mobility protocols; MIPv6 and its extensions are not suitable for these resource constrained devices. In this paper our focus is to study PMIPv6 based mobility management and different Scenarios based on it along with sensor devices. Existing research has made many improvements in terms of HO latency but less attention has paid towards signaling cost and packet loss particularly in time critical areas. The study provides the complete survey of network based mobility management schemes, 6LoWPAN mobility, challenges associated with them and solutions to meet these challenges.

# 1. Introduction

IP Mobility and IOT; one of the current research topics are getting popularity in mobile communication and networked devices. With the tremendous and rapid increase of mobile devices/gadgets connected to the network has given rise to the new field of research for the researchers. Silicon India Magazine recently reported that International Telecommunication Union (ITU) has made an announcement at Mobile World Congress that by the year 2014 there will be a rise from 6 billion to 7.3 billion mobile users and that will be more than the world's population. The rapid growth of Networked devices has led to an anticipated depletion of addresses in the current Internet Protocol version 4 (IPv4), so there is a newer version of IP i.e. IP version 6 (IPv6) [57] which provides sufficient address space to meet the expected increase of network devices. With the introduction of IPv6 there is emergence of new technology called Internet of Things (IOT) [23], where most of the physical objects will be connected to the network. In IOT [23] most of the objects are mobile and therefore requires a mobility management protocol for maintaining IP mobility. Also the users carrying mobile gadgets like cell phones, laptops or smartphones etc. want to remain connected to the network services all the time while they are on the move (i.e. moving from one network to another). In order to provide the mobile users uninterrupted services or to track the signaling from moving objects, there is need of mobility management schemes. In IOT (machine to machine (M2M) communication) [58], IPv6 addressed Sensor nodes (SNs) are being used for making the objects to communicate. IPv6 over Low-Power Personal Area Networks (6LoWPAN) standard allows these heavily resource constrained SNs to connect to IPv6 networks. Mobility in 6LoWPAN is based on standardized network mobility management protocols [11]. Different researchers have used 6LoWPAN mobility with different perspective and application areas like military surveillance, environmental monitoring, healthcare monitoring, vehicular networks etc. where time criticality of data is most important. Mobility management schemes have come over time with the growth of Mobile devices. Some uses mobile hosts to carry mobility while others use network components for the same. Therefore these are classified into two different categories: Host Based Mobility Management schemes and Network Based Mobility Management Schemes. In this context this paper further presents a recent survey of mobility management schemes with emphasis on network based 6LoWPAN mobility management.

**1.1 Host Based Mobility Management Schemes:** In such Mobility Management Scheme, the Mobile host/Mobile Node (MH/MN) which experiences the mobility from one network to another is involved in all signaling related process which requires protocol stack modification and IP address changes on the MN for the session continuity during handover. This signaling process includes movement detection, Router Solicitation request (RtSolReq), Duplicate address detection (dad) and Binding updates (BUs) etc. we have well known Host Based Mobility Schemes which have been used for different applications including real time services and these are: Mobile IPv6, Hierarchical MIPv6 (HMIPv6) and Fast Handovers for MIPv6 (FMIPv6) [4, 5, 7].

**1.1.1 Mobile IPv6:** Mobile IPv6 (MIPv6) [1, 2, 6, 7, 9, 13] allows nodes to move between different subnets, while maintaining reachability and on-going connections between mobile and correspondent nodes. To do this, every time a mobile node moves across subnets, it sends Binding Updates (BU) to its Home Agent (HA) and all Correspondent Nodes (CN) it communicates with, in order to inform them the new IP address, i.e. the Care Of Address (CoA). MIPv6 also uses the Route Optimization mechanism as a default to eliminate the Triangle Routing problem. By Route Optimization, the mobile node and correspondent node can communicate directly without passing packets via the Home Agent (See Fig 1).

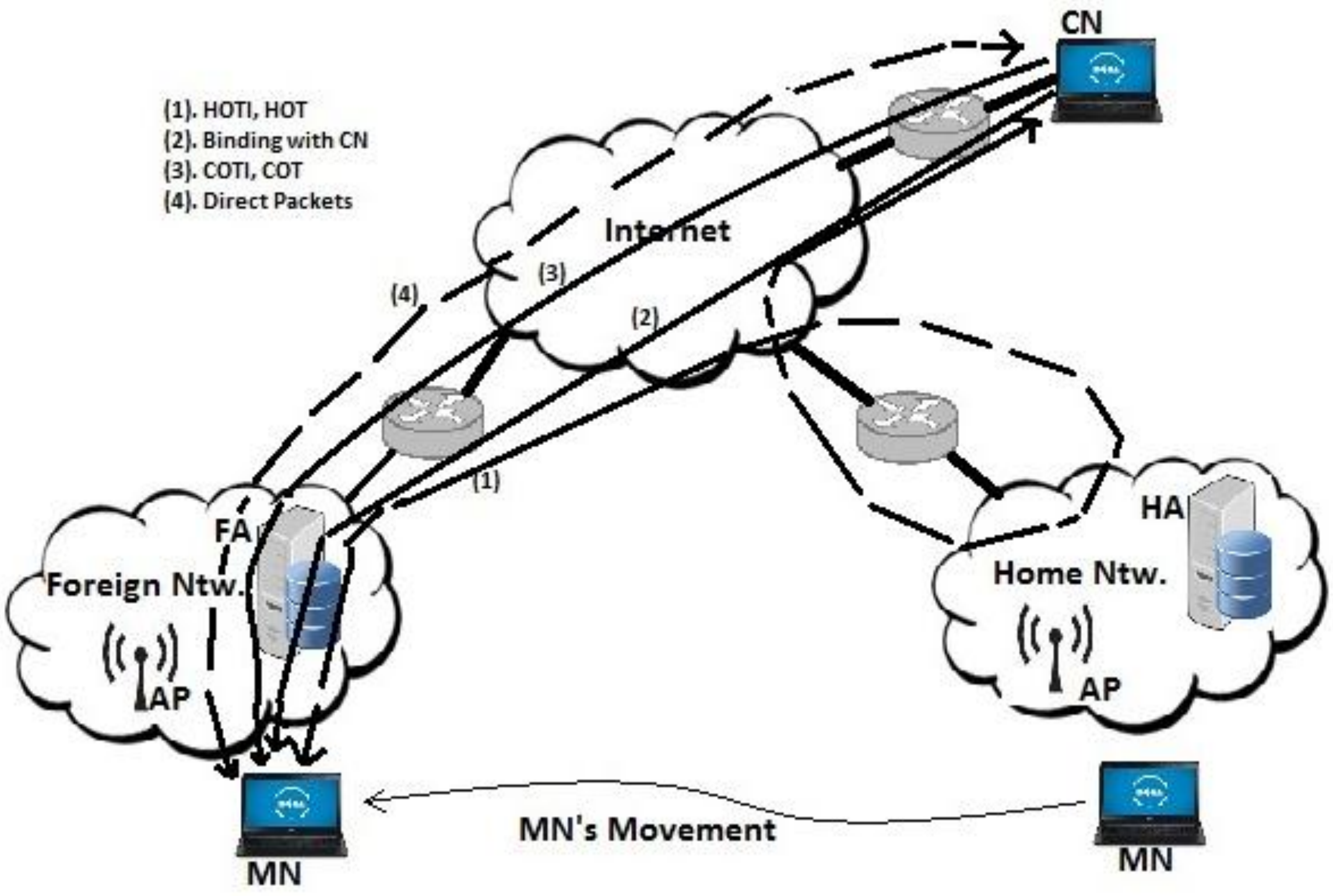

Fig. 1: MIPv6 Basic Scenario with route optimization

**1.1.2 Hierarchical MIPv6 (HMIPv6):** HMIPv6 [3, 4] introduces the hierarchical mobility management model as shown in the figure 2. HMIPv6 works on the same approach to take care with the movement in the region (that is, micro-mobile or local mobility) and inter domain Mobile (macro mobile or global mobility). HMIPv6 [2] introduces mobility anchor point (MAP) so that the movement in the region and inter-domain region can be distinguished. Mobile Node (MN) gets registered with MAP by sending binding updates (BU) to the MAP, thus MAP acts as home agent for MN. Then all the packets for MN are intercepted by MAP and re-transmitted to MN. MN is having two addresses when it is in the MAP's domain: the regional care-of address (RCoA) and local/Link care of address (LCoA). The BU sent to Map by MN consists of RCoA and LCoA. This makes MN's registration of RCoA with the HA and CN. When MN only moves in the MAP region, it only needs to register a new LCoA to MAP and there is no need to send BU to HA and CN because RCoA has not changed thus it reduces the signaling overload over the network and also reduces the packet loss and ultimately reduces the handover delay (See figure 2).

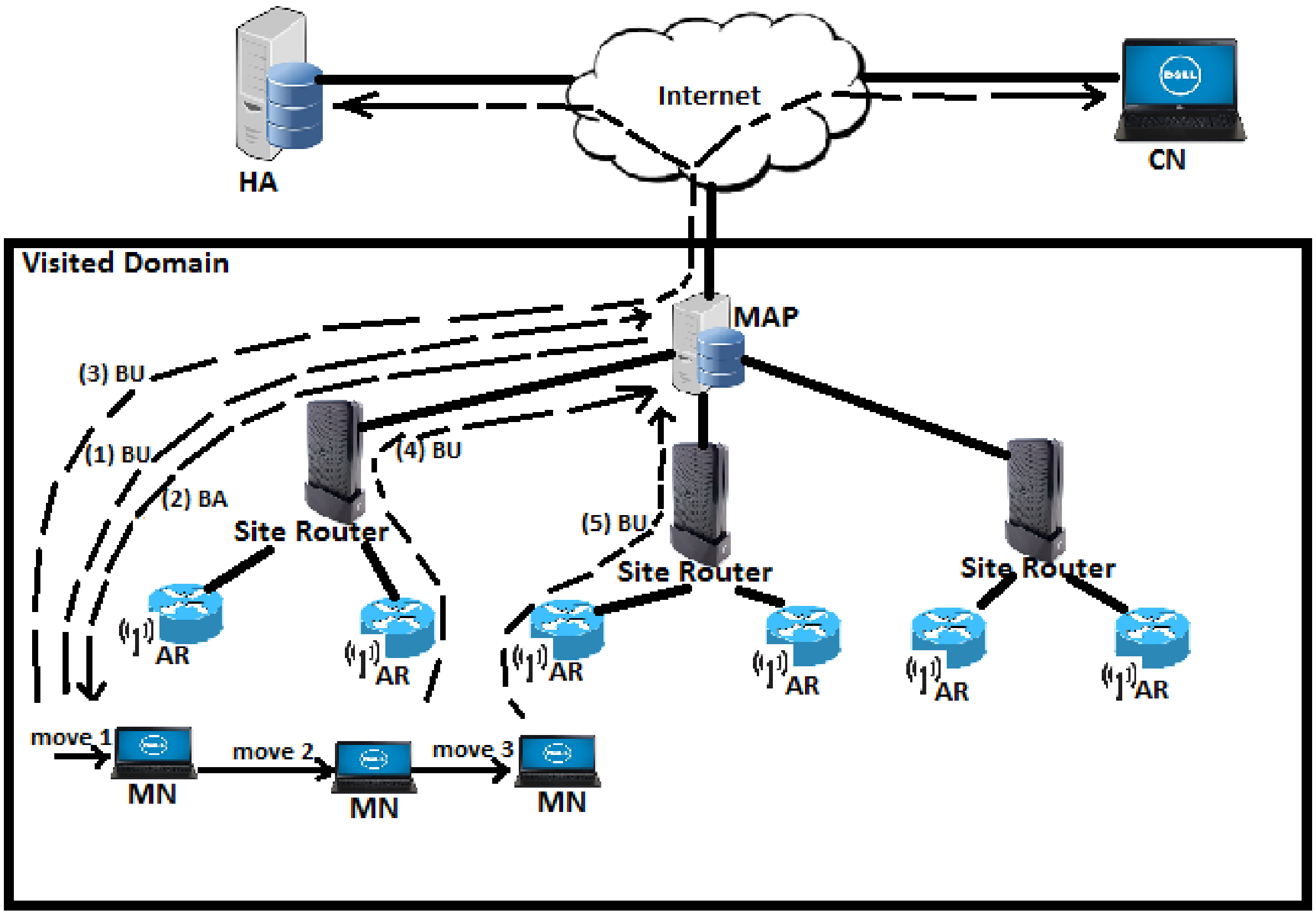

Fig. 2: HMIPv6

**1.1.3 Fast Handovers for Mobile IPv6:** Fast Handovers for Mobile IPv6 (FMIPv6) [5, 3, 7, 8] is implemented based on the MIPv6. Its goal is to reduce the handover latency. The principle is to let the mobile node establish a new temporary address with the new access router before breaking connection with the previous access router. For implementation of FMIPv6 over IEEE 802.11 wireless network, RFC4260 divides FMIPv6 mechanism into two main modes: Predictive mode and Reactive mode.

**1.1.3.1 Predictive Fast Handover Mode:** As illustrated in Figure 3, in this mode when the mobile node (MN) realizes that the handoff is necessary, it performs the scan sometime earlier to the handover, and sends a Router Solicitation for Proxy (RtSolPr) message in order to find neighbor access routers. The currently default access router or previous access router (PAR) responds to MN with a Proxy Router Advertisement (PrRtAdv) resolving the specified access point (AP) identifiers. Therefore, it is able to send the Fast Binding Update (FBU) and Handover Initiate (HI) prior to the new access router (NAR) via PAR. NAR confirms the message by sending back a Handover Acknowledge (HAck). Then, a Fast Binding Update Acknowledgement (FBack) with the CoA will be sent from PAR to both MN and NAR. Packets sent to the MN during its handover will be buffered at the NAR. After the handover process finished, only the fast neighbor advertisement (FNA) is sent; and the buffered packet will be delivered to the MN.

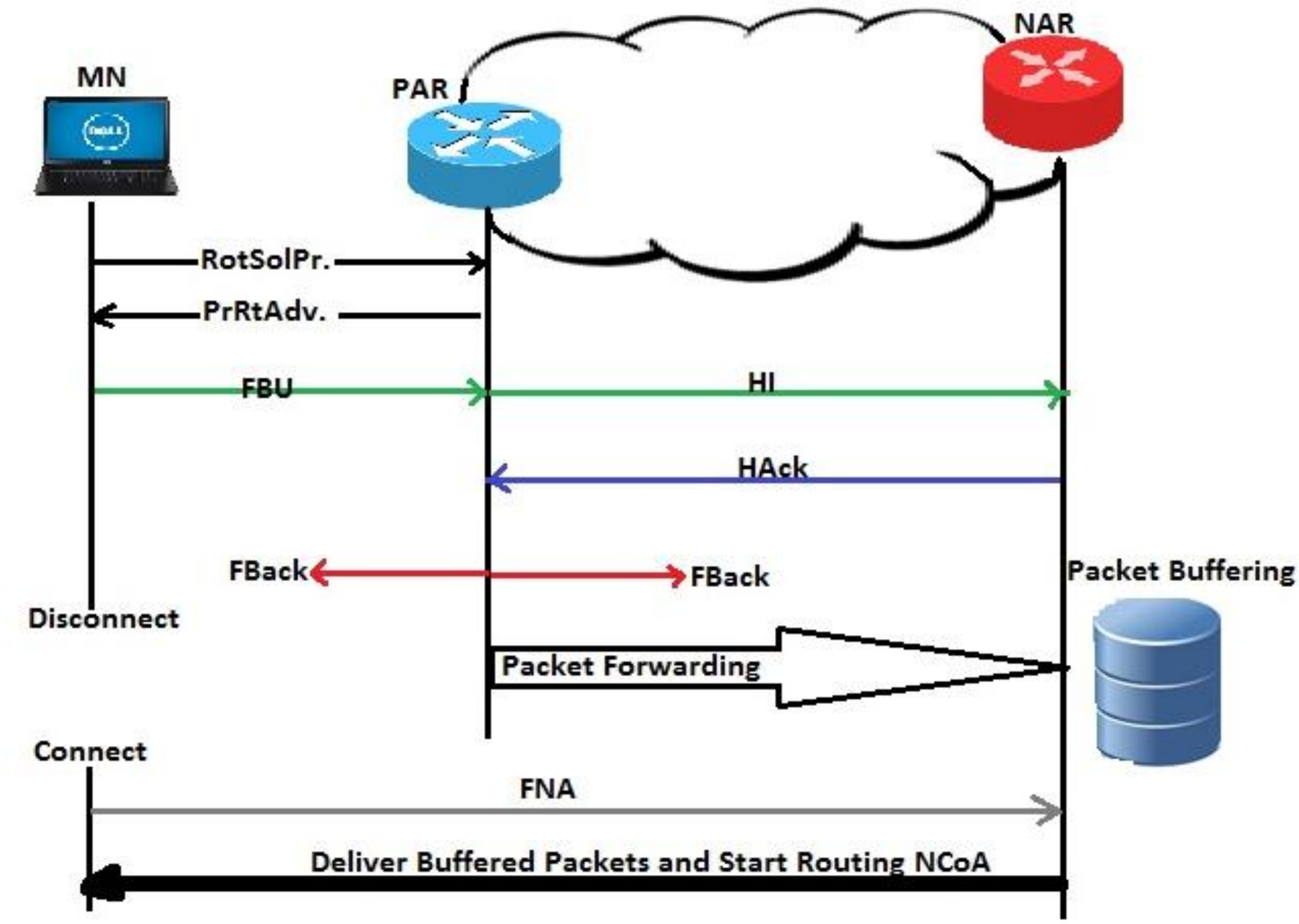

Fig. 3: Predictive Fast HO operations

**1.1.3.2 Reactive Fast Handover Mode:** Mechanism of FMIPv6 in reactive mode is illustrated in Figure 4. Contrary to predictive mode, the wireless mobile node cannot send FBU prior to the handover. During the handover time, packets destined to the wireless node will be buffered at the PAR. Therefore, FBU is sent to the PAR after the handover to inform the PAR to forward packets to the wireless mobile node under NAR (see figure 4).

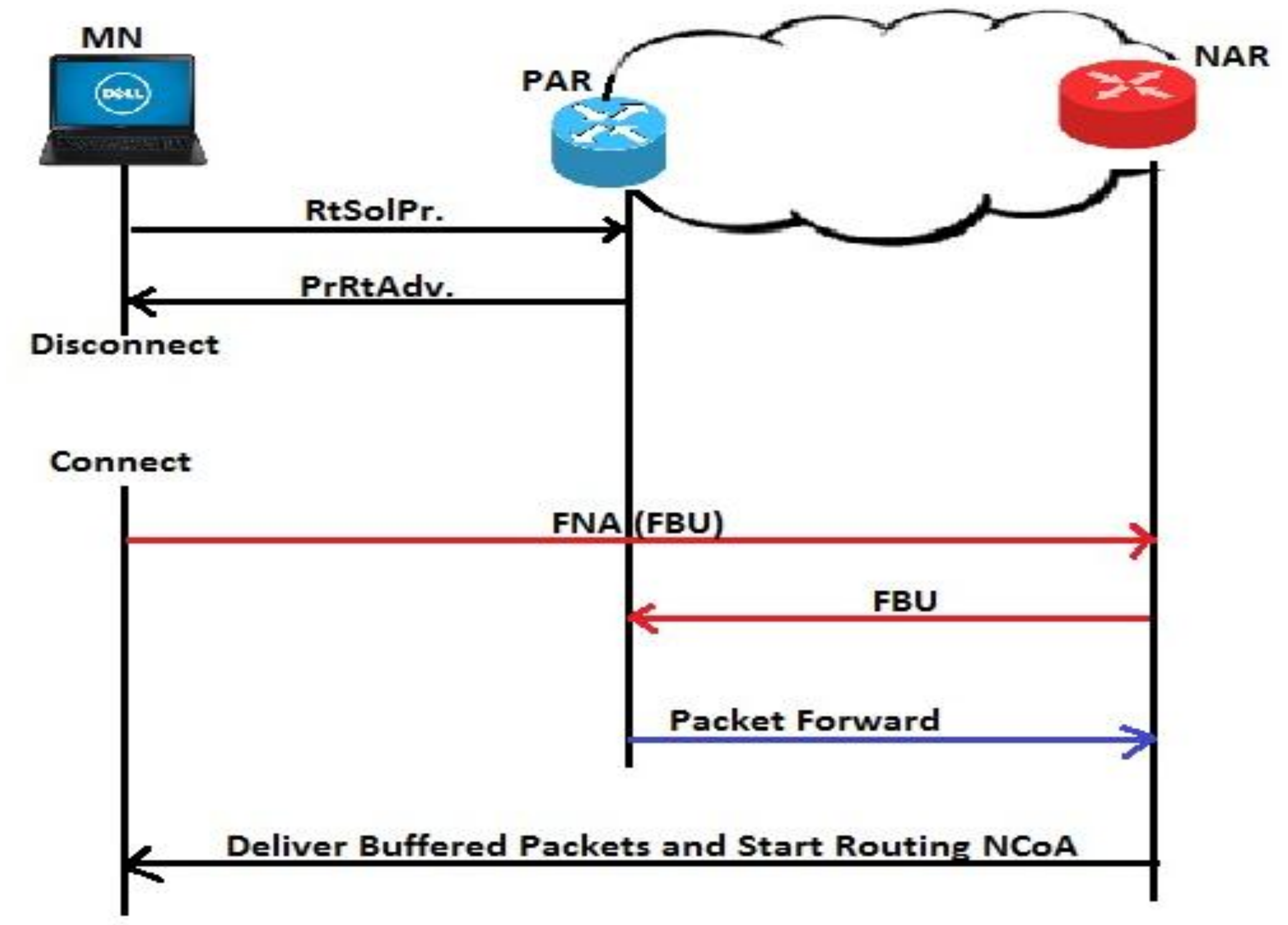

Fig. 4: Reactive Fast HO operations

After the study of Host based mobility schemes, it is clear that all the host-based mobility management protocols basically require protocol stack modification of the MN and the change of its IP address in order to support its mobility roaming within or across network domains. Furthermore, MNs are typically resource restricted (e.g., memory, power) particularly sensor nodes, their involvement in the mobility signaling process (movement detection, Router Solicitation request (RtSolReq), Duplicate address detection (dad) and Binding updates (BUs)) may increase their complexity, consumes power, and wastes the resources. Therefore, these issues and the left over drawbacks, such as high handover latency, packet loss and signaling overhead cost, when put together means that these protocols have not been extensively deployed yet. Kong et al., [32], said that due to their inability to satisfy the quality of service (QoS) requirements for real-time, non-real-time and streaming sensitive services, such as VoIP, video conferencing, audio/video streaming, some improvements are still needed to address these problems.

On contrary, Network Based Mobility Management Schemes attract attention in the Internet and telecommunication societies by improving the performance of the MN's communication to fulfill the requirements of QoS for real-time services. Lot of research is being done on the network based mobility management protocols. Figure (5) below shows the graph related to the ongoing research being carried out on network based mobility management schemes for last 10 years.

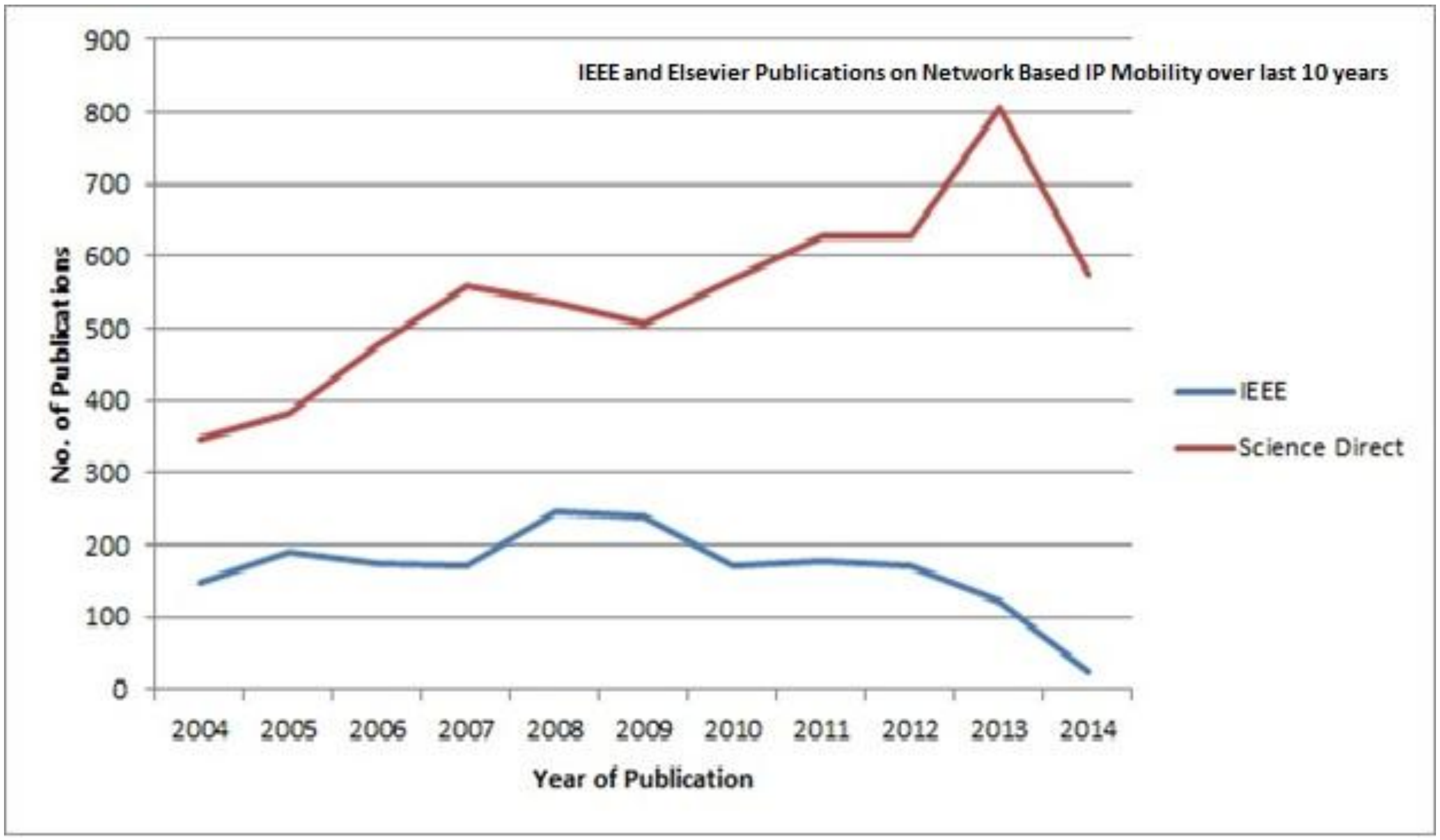

Fig. (5): Number of Publications over last 10 years. Results obtained by submitting query "Network Based IP Mobility" from IEEE (http://www.ieeexplore.ieee.org) and Elsevier (http://www.sciencedirect.com) websites.

Also the standardized protocols meant for network based mobility are found to be very useful for the IP mobility in WSNs and 6LoWPAN. Extensive research is being done on the 6LoWPAN mobility from last few years. The standardized protocols for network mobility have been used for 6LoWPAN and very much satisfactory results are obtained. Below in the figures (6) and (7) are the graphs, indicating research being carried out on IP mobility in WSNs and 6LoWPAN respectively from last couple of years.

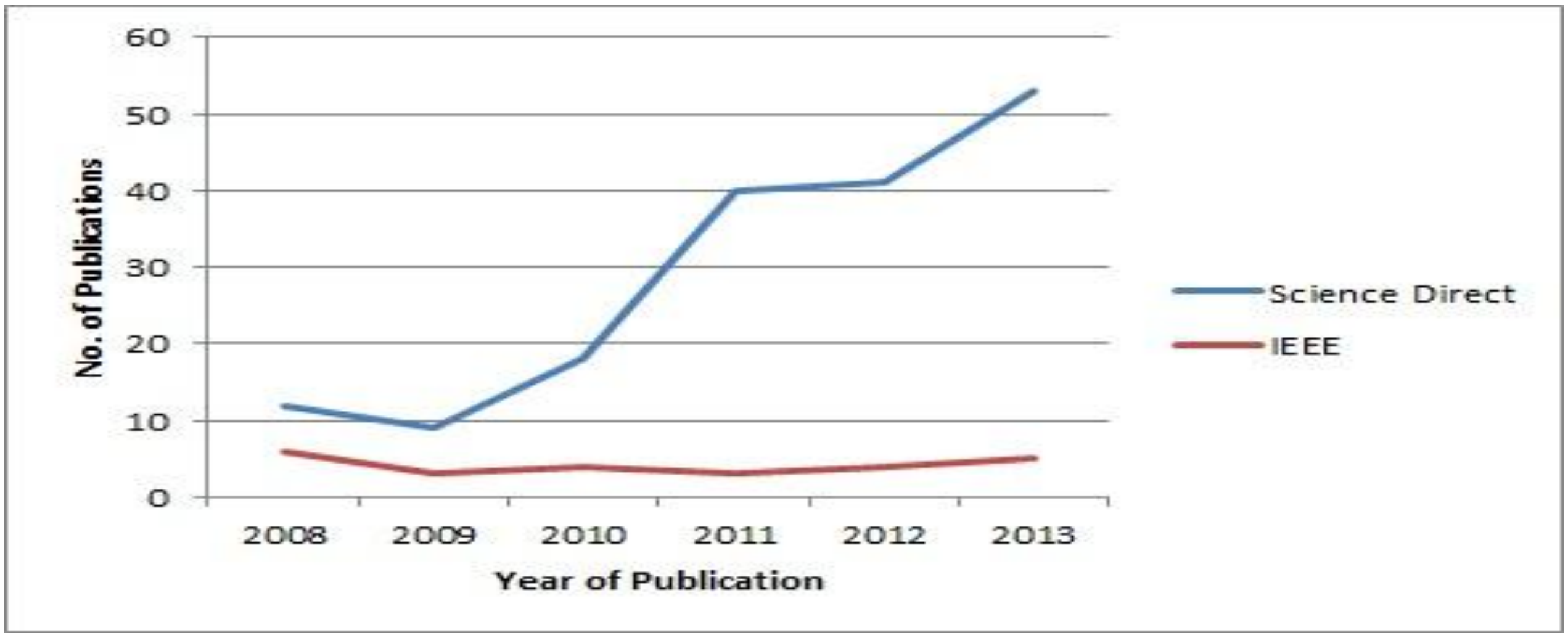

Fig. (6): Number of Publications over last 6 years. Results obtained by submitting query "IP mobility in WSN" from IEEE (http://www.ieeexplore.ieee.org) and Elsevier (http://www.sciencedirect.com) websites.

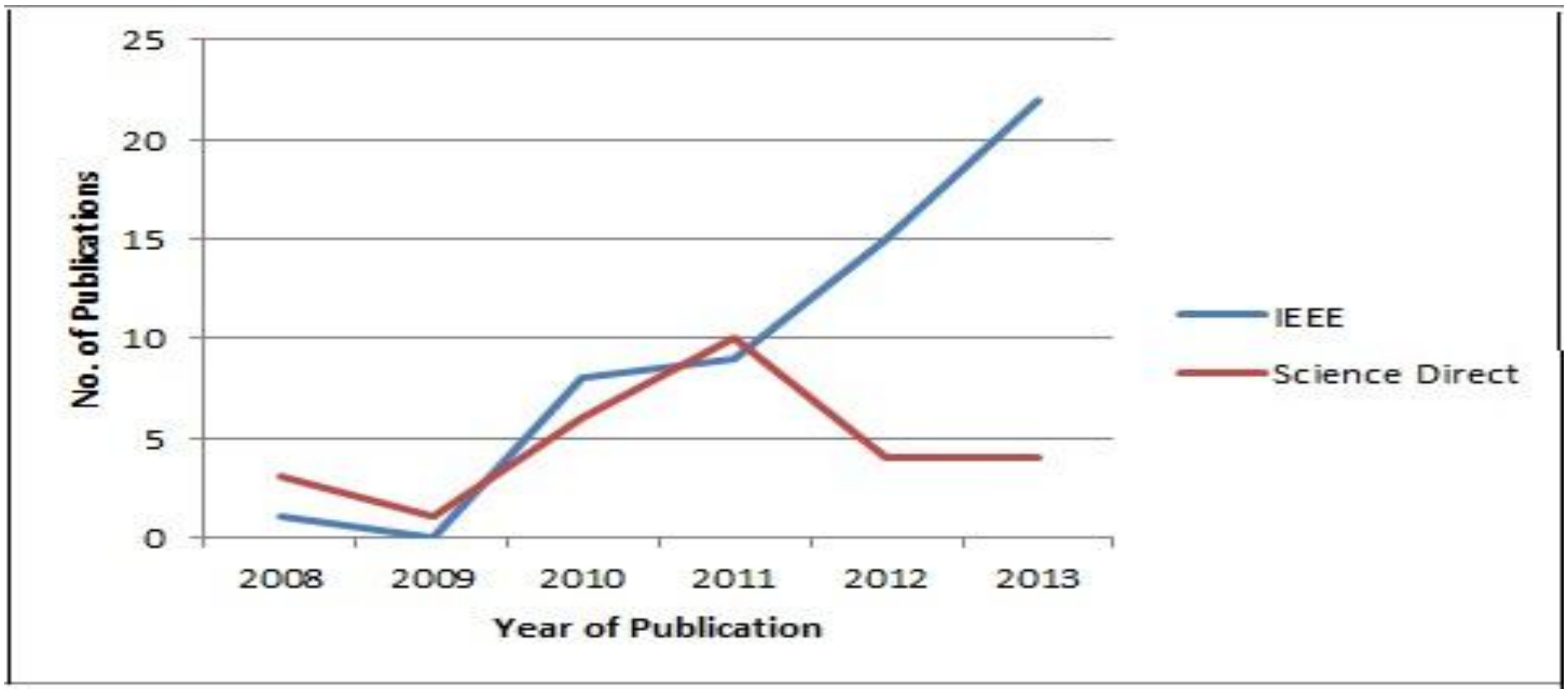

Fig. (7): Number of Publications over last 6 years. Results obtained by submitting query "IP mobility in 6LoWPAN" from IEEE (http://www.ieeexplore.ieee.org) and Elsevier (http://www.sciencedirect.com) websites.

**6LoWPAN:** wireless sensors are being actively used in mission critical applications such as battle field control, medical assistance and natural disaster forecasting. Sensors have distinguishing characteristic of limited resources (memory and processing) and autonomous but with limited power supply. ZigBee 802.15.4 protocol stack [61], a non-IP protocol used for Wireless Sensor Networks. Zigbee has found incompatible with IP and introduces many other constraints like resource usuage, limited bandwidth, energy consumption etc. Due to the urgent need of connecting WSN and internet, a new IETF working group namely 6LoWPAN [62] was established. 6LoWPAN working group also produced two RFCs as RFC 4919 [62] and RFC 4944 [63]. The maximum packet size for transmission in IEEE 802.15.4 [64] is 127 bytes as the sensors are not capable of holding complete IPv6 address. The limited packet size maintains the low power consumption in sensor devices. The 6LoWPAN protocol stack is given in the figure 8.

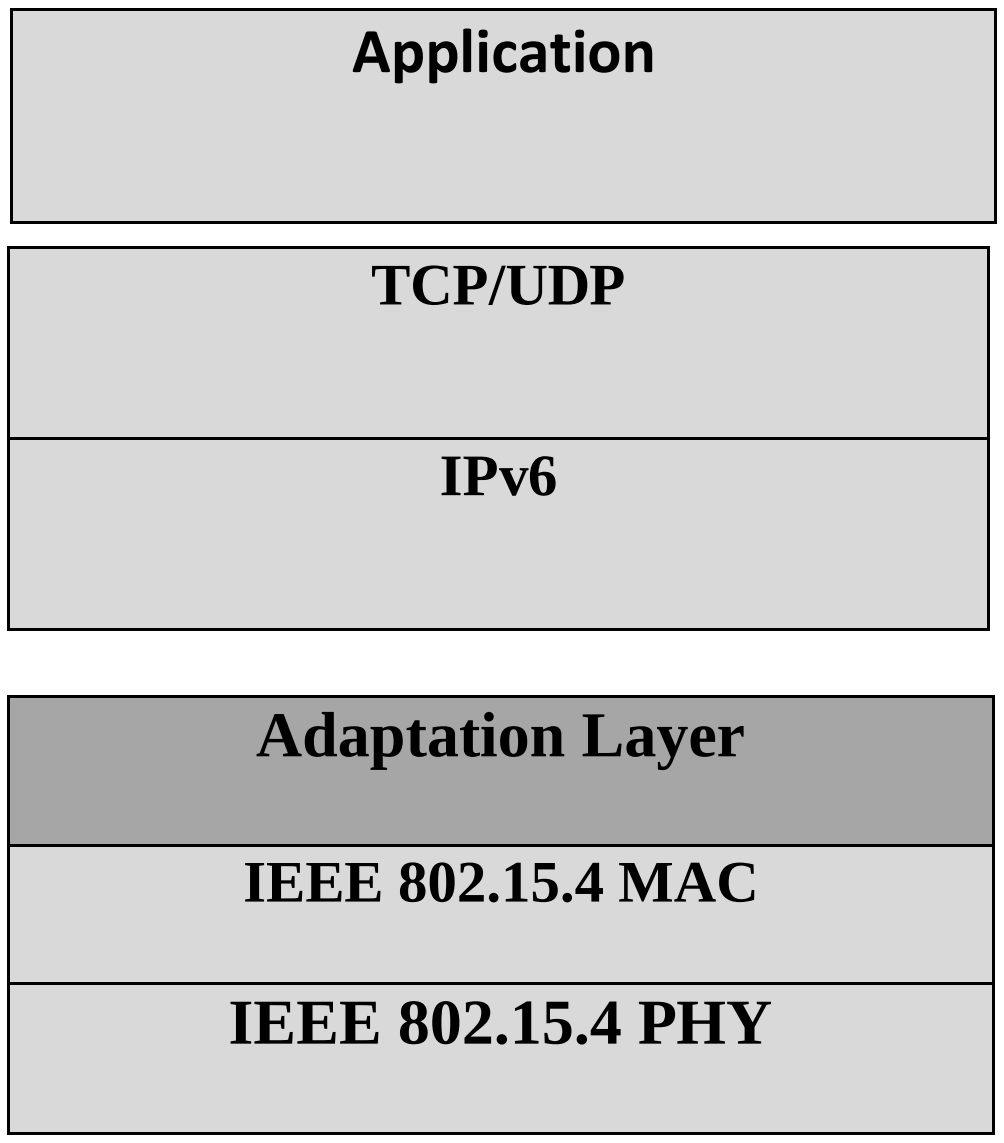

Fig. 8: 6LoWPAN protocol stack

6LoWPAN adopts the PHY and MAC layer protocols of IEEE 802.15.4 and along with IPv6 protocol; it is used as network layer protocol in 6LoWPAN. The adaptation layer provides

seamless transmission between MAC and network layer. Adaptation layer is responsible for fragmentation, reassembly of fragments and IPv6 header compression.

## 1.2 Network Based Mobility Management Schemes:

In Network Based Mobility Management Schemes there is no involvement or less involvement of MN in signaling process (Network components are involved in Signaling). Following protocols were produced over time by the IETF working group and lot of work has been done on these Network Based Mobility Management Protocols.

### 1.2.1 NEMO-BS

**NEMO-BS**: The NEMO [37, 59] Basic Support Protocol has been standardized for IPv6 [RFC3963] but drafted for IPv4 [Leung06] [12]. NEMO is an extension of Mobile IP that facilitates the whole network to change its attachment point to the Internet. Under NEMO, a Mobile Router (MR) takes over the role of the MN in carrying out mobility functions. Nodes that are attached to a MR are called Mobile Network Nodes (MNNs), are not aware of the network's mobility and do not implement any mobility functions. MRs also sends binding updates to their home agents (HAs). However, binding updates from MRs also contain the mobile network's network prefix (MNP). HAs will bind an entire network prefix to the MR's CoA and forward all packets for that network (prefix) to the MR.

Figure 9, shows the path of packets using NEMO [12, 37]. IP packets from a correspondent node (CN) that are destined for a node on a mobile network (MN) are delivered via standard routing on the Internet to the HA of that MN. The HA tunnels the packets to the MR for delivery to the MNN. Reverse packets take the same path in the opposite direction; the MNN send packets to the MR to be tunneled to the home agent and then sent out to the CN via standard routing on the Internet.

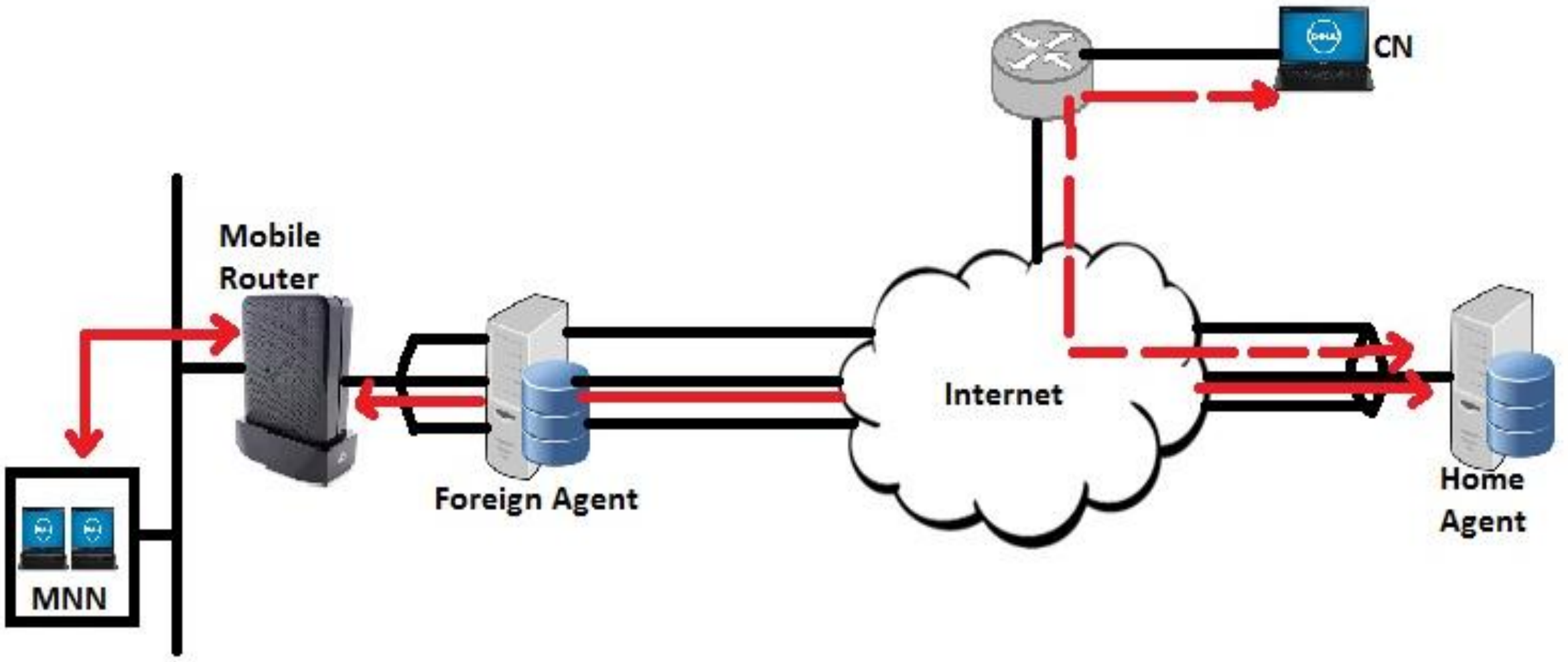

Fig. 9: IP traffic between a Mobile Network Node and a Correspondent Node using NEMO

**1.2.2 PMIPv6:** Proxy Mobile IPv6 (PMIPv6) [11, 60] is a network-based mobility management protocol, the only standardized network based mobility protocol by IETF, specified in RFC 5213 [11]. This protocol used for building a common and access technology independent of mobile core networks, accommodating several access technologies such as WiMAX, 3GPP, 3GPP2 and WLAN based access architectures.

## 2. Work Carried Out over Network Based Mobility Management Schemes and 6LoWPAN WSN Mobility:

Akyildiz et al. [14], Presented a survey on some mobility management support protocols and gave their comparison. Also proposed a mobility solution at data link layer, network layer, and cross-layer. Their mechanism was focused at heterogeneous wireless network’s mobility, for that they introduced a Network architecture as well. Kempf [15] studied the existing mobility management schemes and offered their shortcoming as the involvement of Host in mobility management. Then he presented a need for network based local mobility management. He also presented the classification of the existing solutions for mobility in localized domain in two sets as Interoperable IP-level protocols and Link specific or proprietary protocols. Then he stated that mobility management schemes are not widely deployed for Interoperable IP-level yet. He further specified the problems that: Even if the small modifications of host stack, limits the broad usage of protocol. The existing solutions only depend on MIP or its derivatives and there is no support in existing solutions for both IPv4 and IPv6. Further in the article, he showed few limitations of Link specific or proprietary protocols that handle localized mobility. To address these limitations he suggested PMIPv6 as a localized mobility management protocol to reduce the signaling update time and shorten the disruption period. Therefore, PMIPv6 handover can be comparatively faster than the MIPv6 by using the link layer attachment information. S. Gundavelli et al., [11]**,** Introduced Proxy Mobile IPv6 (PMIPv6) which is intensely standardized by the IETF-NETLMM working group (2008). They presented it as a network-based localized mobility management protocol. This protocol was expected to support the real deployment of IP mobility management where Mobile Nodes (MNs) were not supposed to carry the signaling related to mobility. Because of this significant characteristic of PMIPv6, which carries the mobility management on behalf of the MN without its participation in the mobility-related signaling, this mobility management Scheme is being actively used in many mobility required applications. M. Shin et al., [43, 44] described PMIPv6 network based mobility protocol as a solution to handle the mobility of body sensors. Mobility related signaling is performed by the network side and no need of duplicate address detection (DAD) is required. The control information is exchanged only between Local Mobility Anchor (LMA) point and Mobile Access Gateway (MAG) to reduce signaling cost and handoff latency of each sensor node (SN).

Ping Dong et al., [28] Proposed an Identifiers Separating and Mapping Scheme (ISMS) as a network-based mobility management scheme that takes advantage of the identity and location separation. It satisfies the requirements of faster handover, route optimization, advanced management, location privacy and security. They showed average handover delay of ISMS is on the order of ms only, which is far smaller than that of Mobile IPv6. Also the proposed mechanism can reduce packet overhead on wireless channels to avoid delays. Bag et al. [45] presented a 6LowPAN mobility supporting scheme dependent on dispatch types of the 6LoWPAN. This scheme was to reduce the packet loss and handoff delay but it was meant for Intra domain mobility only and is not suitable for inter-domain problems. Minkeun Ha et al. [39] proposed fast and seamless mobility protocol to support inter PAN (inter domain) handover (HO) in 6LoWPAN and named that method as Inter Mario. They used a SN as partner node to act as access point (AP) which preconfigures the future HO (Predicted by the movement detection algorithm by using Signal Strength indicator) for MN. The partner nodes sends the pre-configuration message (MN Identifier, HoA, The address of HA) to the new Gateway (GW) of the neighbor PAN. The new coordinator in the neighbor PAN notifies the foreign Agent (FA) with the identifier of new comer. FA checks whether it is preconfigured and then sends surrogate BU [MN’s home address.> FA] to home agent (HA). In this way they showed the fast and seamless handover by using MN pre-configuration approach (based on make before break) even before the handover actually occurs. Their scheme is found efficient for reducing packet loss in inter-domain mobility. Juha Petajajarvi and Heikki Karvonen [40] proposed a soft handover method for mobile WSN. In their work they took results over testbed for mobile and static gateways along with mobile Sensor Nodes (SN). Also an algorithm for connection quality check was presented to eliminate unnecessary handovers and a comparison of Proposed SH-WSN6, MIH-PMIPv6, FPMIPv6 and PMIPv6 was done by considering the metrics Wireless link delay, Router distance latency, Delay between LMA and nMAG Vs. Handover latency but their connection quality algorithm lags in providing optimal value for threshold to perform handover. L.M.L. Oliveiri et al. [46] Presented Routing and mobility approaches for 6LoWPAN mesh networks but the requirements and resources in the existing solutions to 6LoWPAN for their adaptation is still a challenge and further research on 6LoWPAN mobility is required. Sergio Gonzalez et al. [49] Proposed an approach supporting health monitoring mobility at home by using wearable body sensors. One body sensor acts as coordinating node, makes communication between rest of body sensors and APs. In their proposed handover procedure, the coordinator node utilizing all the 16 channels of IEEE 802.15.4, sends a PING message. In its acknowledgement message from APs, the coordinator gets the RSSI value of all surrounding APs. The AP having best RSSI value is chosen and health related data is sent to the network through this AP. If the RSSI value is decreased below some predefined threshold, then the body sensor at that time having better verified RSSI value acts as temporary coordinator to carry the process. They did not clearly mention the body sensors are FFDs or RFDs because to behave like a coordinator, SNs must have router functionalities. Myung-Kyu Yi et al. [10], Studied performance of PMIPv6 and compare it with that of hierarchical Mobile IPv6 (HMIPv6). They proposed an investigative mobility model based on the random walk to take into account several mobility situations. They also articulated the location management cost and handoff management cost. Further evaluated the performance of the proxy mobile IPv6 and hierarchical mobile IPv6, respectively. Although their numerical results show that PMIPv6 can have superior performance to HMIPv6 by reducing the IP latencies for location update and handoff but more improvements are still needed in reducing packet loss. Rong Chai et al. [18] proposed a new network architecture to integrate NEMO and 6LoWPAN-based WSNs to evaluate the performance of NEMO and MIPv6 and also studied about the energy constraints of Sensor Nodes. In their simulation study, they varied number of MNs from 5-25 and considered it as group management scenario. They assumed the total energy of Sensor Router (SR) and Sensor Node (SN) is 10 and through simulation they showed that the remaining energy is the function of simulation time [18]. The Simulation results for MIPv6 and NEMO were shown; they obtained energy depletion of SN for NEMO in comparison to MIPv6 Scheme decreases about 35%. Also they observed the application of NEMO protocol in 6LoWPAN reduces the handoff latency in comparison to MIPv6. Md. Fotouhi et al. [47] proposed mobile sensor node's handover procedure in Wireless Sensor Networks. The metrics used for the evaluation of handover procedure are Reduced Signal Strength (RSSI), Velocity of Mobile Sensor Node, Number of Hops between MN and AP, traffic load, energy level and link quality etc. To know about the link quality, they used Fuzzy Link Quality Estimator (F-LQE) [48]. In their work, first the need for handover was evaluated based on the metrics RSSI and velocity of Sensor MN. Secondly in the procedure, MN Sends continuous regular probe messages to all the surrounding APs and the best AP based on the value of RSSI is chosen and MN gets an acknowledgement from the selected AP. Now, not only RSSI value is taken into account but also the other parameters such as traffic load, depth (no. of hops) and energy level were considered. To choose the best AP to register with, they used F-LQE. Motaharul Islam et al. [22] proposed SPMIPv6; Sensor Proxy Mobile IPv6, a mobility supported IP-WSN protocol which was based on PMIPv6. They presented the architecture of the proposed system, message formats and also analyzed its performance by considering the parameters like signaling cost and mobility cost. The analyses showed that the proposed scheme reduces the signaling cost by 67% and 60% as well as reduces mobility cost by 55% and 60% with comparison to MIPv6 and PMIPv6 respectively. Also by increasing the number of WS-Nodes, Signaling cost increases and increased number of hops increases the Mobility cost. Dizhi Zhou et al. [33] did the theoretical analysis and evaluated the handover latency of PMIPv6, Fast handovers for PMIPv6 (FPMIPv6) and Transient Binding for PMIPv6 (TPMIPv6) through simulation study in vertical handover environment. In addition to this, packets loss rate of UDP traffic and the throughput of TCP traffic of these protocols was evaluated. For handover latency comparison they clearly explained the various factors affecting Handover Delay including: Impact of wireless link2 (new interface) delay, Impact of delay between MAG and LMA and Impact of residence time. Results for handover latency comparison showed that: the residence time has a great impact on handover latencies of TPMIPv6-Predictive and FPMIPv6. By keeping the residence time at 500, 1000 and 1500 (ms), simulations for handover latency, UDP packet loss rate and Declining TCP throughput were carried out. They clearly showed that the handover latency of FPMIPv6 is 1.5 times larger than PMIPv6 at 500 residence time (RT), 1.96 times more at 1000 ms RT and 2.37 times more at 1500 ms RT in predictive mode and it 1.14 times more than PMIPv6 at all different RT in reactive mode, while TPMIPv6 has 0.76 times more HO latency than PMIPv6 at 500 ms RT, 0.675 times more HO latency at 1000 ms than PMIPv6 and same HO latency as PMIPv6 at 1500 ms RT. The UDP packet loss rate of FPMIPv6 in predictive mode is 4% more than PMIPv6 at all three residence times, it 19.4%,19.6% and 19.3% more than PMIPv6 at three different RTs respectively and it is 60%, 25%, 4% more in case of TPMIPv6 than PMIPv6 at the three RTs. Long-Sheng Li et al. [16] tried to improve the vulnerability in security of IP mobility management schemes by suggesting a nested IPsec Encapsulating security payload (ESP) from MN to CN in nested NEMO. They considered the traffic from HA to CN. They have used one-tier IPsec ESP between MN and CN, and assumed the MRs and MR_HAs having the functionality with IPsec. They used ISAKMP for key management in NEMO. The performance result analysis directed that as increasing more levels of nested NEMO, the more it will be secured. In their results they gave a comparison of ESP time and Delay time for Basic NEMO and Proposed system. Also the comparison of packet size was shown for both the schemes; the NEMO and the proposed scheme and results clearly showed some increased percentage for the proposed system. E. A. M. Avelar et al. [30], Described network based mobility management and evaluated a testbed for PMIPv6. The performance evaluation was carried out by considering QoS (Quality of Service) and QoE (Quality of Experience) metrics to see the PMIPv6 support for multimedia traffic and found better results for PMIPv6. A. J. Jara et al. [50] proposed an approach for Wireless Personal Area Networks (WPAN), based on mobile-IP in some critical environments as an extension to IEEE 802.15.4 and named it as GinMAC. GinMAC also provides Intra-PAN mobility support in WSN. Zinon et al. [51] used the GinMAC extension given in [50] to monitor some parameters particularly RSSI for maintaining good link quality to all MNs. RSSI anticipates the MN's movement and direction. MNs can be in communicating or silent state. In silent mode, keep-alive or node-alive approach is used [52]. The registered APs send periodic keep-alive messages and thus the MN knows the exact time to get the keep-alive message. If MN does not get any message then MN sends node-alive message and waits for acknowledgement. If no response is found then MN goes in scanning mode for a new AP and starts the procedure again. Ibrahim Al-Surmi et al., [31], Presented IPv6 features to support mobile systems and survey on the mobility management services along with their techniques, strategies and protocol categories, and explained the categorization and comparison between several mobility management protocols. They also acknowledged and debated on some issues and challenges facing mobility management like QoS, Fault Tolerance, Scalability, Signaling Overhead, Bandwidth Constraints, Cross Layering Issue, Triangular Routing, Security, Handover Latency, Packet Loss, Global and Local Mobility Interaction, Multi-homing Problem and Developing standards for seamless and secure mobility. In addition, there are certain unsolved critical challenging issues, such as the integration of different technologies, interoperability of networks and connectivity, which stops the achievement of complete and continuous mobility within homogeneous and heterogeneous environments. Wang Xiaonan, Zhong Shan and Zhou Rong [36] , Proposed a mobility support scheme for 6LoWPAN. The control information interaction for both the mobility handoff and the tunnel establishment is performed in the link layer, so the transmission unit of the control information is smaller and the delay time taken by transmitting the control information is shorter. The routing process of the control information is automatically performed through the network topology, which saves the delay time taken by establishing the routing paths and reduces the packet loss rate. Also neither does a mobile node need a care-of address, nor does it take part in the mobility handoff control process, which saves the mobile node's energy and prolongs its life span. From the theoretical and simulation studies, they analyzed and compared the performance parameters, including the mobile handoff cost, the mobile handoff delay time and the packet loss rate. The results showed that the performance of the proposed scheme is better than other schemes. Antonio de la Oliva et al. [26] studied basic standards for providing IP mobility support, the functionality attained by combining them and the performance cost of each combination in terms of protocol overhead and handover latency. They identified a strategy for combining mobility protocols and properties that facilitate this combination and have shown that combining different mobility schemes has a non-negligible cost. It is also mentioned that the main contribution to the overall handover time is the layer-2 handover delay and they measured average layer-2 handover delay is about 100 ms. Also highlighted that layer-2 handover has been performed without any optimization and thus a lower delay might be obtained as the movement detection delay/time (mdd/mdt) was proved to be negligible, with an average measured time of less than a millisecond. Jinho Kim et al. [27] defined a protocol for 6LoWPAN mobile sensor node, named 6LoMSN, based on Proxy Mobile IPv6 (PMIPv6). They stated that conventional PMIPv6 standard supports only single-hop networks and cannot be applied to multihop-based 6LoWPAN.They also defined the movement notification of a 6LoMSN in order to support its mobility in multihop-based 6LoWPAN environments. Sofiane Imadali et al. [17], Proposed an IPv6 vehicular platform to integrate eHealth devices that allows the eHealth devices to send the captured health-related data to a Personal Health Record (PHR) application server in the IPv6 Internet. The collected and transmitted data is examined remotely by an expert (doctor) who diagnoses the decease based on that data and provides an immediate decision. They presented a real testbed and an address auto-configuration technique based on a DHCPv6 extension to provide IPv6 connectivity to the resource constrained devices (used for capturing health related data). Ricardo Silva et al. [29], Considered soft and hard handoff by assessing the use MIPv6 and showed that MIPv6 complexity leads to long handoff time and high energy consumption of wireless sensor nodes. To anticipate these problems of MIPv6, a proxy-based mobility approach in which resource-constrained sensor nodes can be relieved from heavy mobility signaling, significantly lessens time and energy overheads during handoff was proposed. The evaluation of both MIPv6 and the proposed solution was done through simulation, by varying the number of nodes, sinks and mobility strategies. A similar approach like in [29] was proposed by R. M. Silva et al., [53]. They used the concept of interconnected proxies having shared backbone, to transfer multimedia data in critical environments in WSNs. In [53], proxy was used to reduce the energy consumption and handoff time. The proxy with best link quality to a MN is chosen as Local Proxy to MN. When local proxy gets deterioration in link quality to MN then other proxies are informed by the chosen proxy. The proxy with better link quality to MN replies back and chosen as next proxy and this is the indication that handoff is performed. Jong-Hyouk Lee et al. [20],

Introduced new NEMO support protocols as P-NEMO, which was based on entities provisioning mobility and were introduced in PMIPv6. In P-NEMO, vehicle mobility management is supported by entities which provision mobility services residing in a given PMIPv6 domain. To further improve handover performance, another efficient protocol FP-NEMO to anticipate the vehicle's handover based on wireless layer 2 (L2) events was proposed. They clearly showed the signaling flow of P-NEMO and FP-NEMO and found the number of message exchange is more in FP-NEMO than P-NEMO but packet loss ratio is lesser. Mun-Suk Kim et al. [34], Presented a thorough analysis to evaluate the performance of PFMIPv6 in terms of the handover latency, the packet loss, and the signaling overhead, in comparison with PMIPv6. The analysis was also validated by simulation study. Results showed that PFMIPv6 improves the handover performance over PMIPv6, especially in the highway scenario where the degree of certainty for an anticipated handover is more, while it performed worse than PMIPv6 for slow mobiles in the city scenario as it takes too long for the slow mobiles to arrive at the nMAG since the predictive handover is triggered. To resolve this problem, they proposed, to perform a handover in the reactive mode for slow mobile in the city environment, although the pMAG receives an L2 report from the mobile, which is referred to as the hybrid scheme. It is shown via analytical and the simulation results that the hybrid scheme achieves shorter handover latency and smaller packet loss than both PMIPv6 and PFMIPv6, while not incurring any additional signaling cost compared to PFMIPv6. They also discussed that the simulations have been performed in the realistic vehicular network configuration to give an insight that the analysis results of which match with the simulation results. Mohammadreza Sahebi et al. [19] described a mobility solution for mobile patient node (MPN) in hospital premises to maintain the continuous connectivity between the patient nodes and hospital area network. For case study hospital architecture was considered to show that their proposed solution reduces the amount of messages exchanged between the MPN and 6LoWPAN hospital network, and also stated that it reduces traffic on Mobile Router (MR). further they showed the comparison of message exchange for three cases; message exchanged with set of Sensor Nodes, Messages exchanged with one MR and Messages exchanged with proposed optimized MR. Finally, it was shown that their scheme provides the same handoff cost and light traffic on MR and Border Router (BR) irrespective of the number of sensors deployed on the patient node's body. Jong-Hyouk Lee et al. [21], Analyzed and compared existing host based IPv6 mobility management protocols like MIPv6, FMIPv6 etc. including the recently standardized network based PMIPv6 and FPMIPv6. Also identified the characteristics of these IPv6 mobility management protocols and evaluated performance by examining handover Latency over Frame Error Rate. Further analyzed the performance of the IPv6 mobility management protocols by considering the performance metrics like handover latency, handover blocking probability, and packet loss over velocity of MN in ms, Frame Error Rate etc. and showed their comparison graphically. For packet loss it follows the sequence as FPMIPv6<PMIPv6<HMIPv6<MIPv6 and for handover latency it follows the sequence as FPMIPv6-Predictive<PMIPv6<FPMIPv6-Reactive<HMIPv6 & FMIPv6 (almost same) <MIPv6 over frame error rate. Julien Montavont et al. [23], Evaluated Mobile IPv6 over 6LoWPAN. They executed Mobile IPv6 in the Contiki operating system and accomplished thorough experimentations on a real testbed. They also proposed a new mechanism called Mobinet, for movement detection. Mobinet was based on passive overhearing. Their results highlighted that Mobile IPv6 can be a useful solution to achieve layer 3 mobility on 6LoWPAN. The layer 3 handover only takes 1.5secs on average with full header compression and is more than satisfactory although it does not permit the support for real-time communications. Prem Nath and Chiranjeev Kumar [24] introduced an AMAP (Adaptive Mobility Anchor Point) to reduce the regional registration cost and packet delivery cost in IPv6 networks. The proposed

AMAP is a special mobility anchor point which was on the basis of activity rate (ARate) of mobile units (MUs) under any AR's domain. AMAP is very valuable in location management of those MUs which follow fixed mobility pattern and they again stated if the HA knows the Mobile Node's daily route (they presented the scenario of a person following same route while going from home to office), the signaling cost due to location update and data packets delivery can be eradicated significantly. They said that HA can store the MU's mobility profile (if fixed mobility pattern of MU) which includes the information related to the list of Access Routers (ARs) traversed, activity rate (ARate) under each AR, enter and exit time under an AR and average speed of MU(or MN) under AR domain. The AMAP is placed on the top of the hierarchy of ARs to lessen the regional registration and packet delivery cost. Their results for location update cost by varying the speed of MU (0-35m/s) showed that the proposed scheme (with AMAP) has less cost as 0-74.11% less, 0-76.4% less and 0-75.6%less in comparison to MIPv6, HMIPv6 and PMIPv6 respectively. Cheng-Wei Lee et al. [25] Proposed 2MR network mobility scheme which takes advantage of the physical size of high-speed trains to deploy two mobile routers (MRs) in the first and last carriages. This scheme provides a protocol to allow the two MRs to cooperate with a wireless network infrastructure in facilitating seamless handovers. The simulation results demonstrate that compared to the traditional single MR schemes, the 2MR scheme noticeably improves the communication quality during handover by significantly reducing handover latency as well as packet loss for high-speed trains. Zinon Zinonos et al. [35], Proposed mobility solution proficiently sustains the connectivity of the mobile node by governing the handoff procedure. The proposed solution was a Fuzzy Logic-based mobility controller to benefit sensor Mobile Nodes (MNs) to decide whether they have to initiate the handoff procedure and perform the handoff to a new connection point or not. In the design of their solution, network state variables which are freely available at all, sensor MNs were used.

The proposed solution is generally applicable to any industrial WSN or testbed with mobility requirements. They validated their proposed mobility solution on a real testbed scenario inside the industrial environment of an oil refinery. The results of the experimentation clearly showed that the proposed mobility solution overtakes the RSSI-based mobility solution in terms of packet loss, packet delivery delay, energy consumption, and ratio of successful handoff triggers. Yuh-Shyan Chen et al. [41] considered group based network roaming in PMIPv6 domain in 6LoWPAN to propose an enhanced existing group based mobility scheme. In their work they overcome the previous existing schemes in [42] and [18] which were relied on the "first newly attaching node in the new domain will carry the rest of node's binding information to reduce the signaling cost" however, sensors on the human body attach to the new access link at the same time. So they presented that it is good to group the body sensor to enhance the procedure and use one (RS & RA) message to carry whole body sensor's information. In addition to this, new Router Solicitation (RS) and Router Advertisement (RA) message formats to combine the necessary information of sensors into one message for reducing signaling cost were provided. Finally they gave a comparison of the original protocol, group based protocol and proposed protocol and stated that their proposed scheme provides better results in terms of signaling cost, average delay time and packet loss ratio.

More literature based on Network mobility and 6LoWPAN mobility is available on different resources of literature. Many researchers have used same protocols but in different perspective and applications. 6LoWPAN mobility is catching pace with the use of standardized network mobility management protocols. Table 1 showing the summary of a comparison between some of the mobility solutions based on network mobility protocols for wireless sensor networks (WSNs).

TABLE 1

Comparison of available mobility solutions based on network mobility for WSNs and 6LoWPAN

| **Author** | **Support of Multi-hop** | **Type of Movement** | **Handoff carried by:** | **Performance metrics used** | **Scheme or Protocol used** | **Cont. Message Exchange** | **Frequent Handover** | **Mobility domain** |
|---|---|---|---|---|---|---|---|---|
| G. Bag et al. [45] | yes | Random | AP's Side | RSSI and movement detection delay | PMIPv6 | yes | yes | Intra-Domain |
| Jara et al. [54] | yes | Random | AP's side | RSSI and movement detection delay | PMIPv6 | yes | yes | Intra-Domain |
| Minkeun Ha et al. [39] | No | Random | GW's side | Signaling cost (SC), End to End delay, Resident time (RT) | MIPv6 | No | No | Inter-Domain |
| Dizhi et al. [33] | No | Random | MAG's side | HO latency, Wireless-link delay, RT | PMIPv6, FPMIPv6, TPMIPv6 | Not Defined | No | Intra-Domain |
| Jian Xie et al. [55] | Not Defined | Random | AR's Side | SC, HO delay, No. of hops | MIPv4, MIPv6 | yes | yes | Inter-Domain |
| Islam et al. [22] | yes | Random | SMAG's side | SC, Mobility Cost (MC) | PMIPv6 | yes | yes | Intra-Domain |
| Long Sheng et al. [16] | yes | Random | Mobile Router's (MR) side | Delay Time, ESP times, Throughput, Jitter | NEMO | Not Defined | Not Defined | Inter-Domain |
| Sergio Gonzalez et al. [49] | No | Random | AP's side | RSSI, Packet Loss | Not defined | No | No | Intra-Domain |
| Juha petajajarvi et al. [40] | No | Random | GW's side | Wireless-link delay, HO delay, Router distance latency, Delay b/w LMA and MAG | SH-PMIPv6 | RA/GW messages are cont. while from SN side depends upon calculated connection quality | No, depends upon connection quality comparison algo. | Not Defined |
| Hana Jang et al. [56] | No | Random | SMAG's side | SC, MC | mSFP, PMIPv6 | yes | yes | Inter-Domain |
| Jong Hyouk Lee et al. [20] | yes | Random | MR's side | SC, HO delay time | P-NEMO, FP-NEMO | yes | No | Intra-Domain |
| Mohammadreza Sahebi et al. [19] | No | Random | Border Router's side (BRs) | HO cost, Traffic overload | Not Defined | yes | yes | Intra-Domain |
| Mun-Suk Kim et al. [34] | No | Random | MAG's side, BS's side | HO latency, SC, Packet Loss | PFMIPv6 | Not Defined | yes | Inter-Domain |
| Cheng Wei Lee et al. [25] | yes | Specific and Random as well | MR's side | HO delay | FPMIPv6 | Not Defined | Not Defined | Inter and Intra Domain |

## 3. Open Issues and Challenges

From the literature survey, we conclude that lot of work has been carried out on IP mobility management schemes to provide seamless handover to MN. Some of the existing methods are host based while others are network based. Network mobility protocols found to be very useful in 6LoWPAN mobility. Such networks are composed of devices with limited energy resources, memory and computational power. Recently, the research community enabled IPv6 connectivity in those networks by the means of an adaptation layer [22, 27]. Lot of research has been done and is being carried out on 6LoWPAN WSN mobility (see figures (6) and (7)). The focus of the research is to reduce signaling cost, packet loss and particularly HO latency. HO latency is caused by L2 and L3 handoffs. Channel scanning, authentication and association delays contribute L2 delays while as movement detection (mdd), CoA, duplicate address detection delay (dad) and registration delay contribute L3 delays [65]. The most time consuming delay is channel scanning and [66, 67, 68, 69] make some improvements in reducing L2 delay. [68] Used Pre-registration to reduce HO delay while [67] used caching AP strategy for the same. The group based protocols in [18, 42] reach the goal of reducing the signaling cost due to carrying of binding information by the newly attaching nodes in WBAN. But sensors in WBAN attach to new link at the same time. Therefore in [41], one control message (RA and RS) to carry the whole body sensor's information was used to reduce the signaling cost. In [70], FPMIPv6 was further used to reduce the signaling cost in sensor networks.

Therefore we realize that Healthcare is one of the research fields, growing rapidly on the basis of these 6LoWPAN WSN. Many healthcare applications use the existing mobility management protocols. Although these schemes are providing acceptable results but still suffer from few shortcomings:

In this article we take the hospital wireless sensor network (HWSN) as case study. Continuous Patient's health monitoring in HWSN is very essential. The patients are autonomous and mobile. To support mobility in HWSN, the signaling messages (RA and multicast messages) used for registration process should be few to enhance the SN's life, deployed on patient's body.

HWSN is controlled network and mobile node (SN) is well aware of the infrastructure. So the continuous exchange of messages between mobile SN and AP to check the signal strength can be avoided which saves life of battery operated devices and enhances their mobility.

Also multi-hop communication adds up extra signaling overhead between SN and AP which also leads to drain of SN's battery. Single-hop communication can be achieved by deploying extra APs, although it will increase expenditure but provides Signaling cost effective and seamless mobility.

In HWSN, sensors on patient's body generate critical data including patient's body parameters such as Pulse rate, ECG, Sugar Level, Body Temperature, Blood pressure etc. This data is time critical and should be transmitted to Hospital Monitoring Station without any packet loss. An expert sitting at monitoring station has to give immediate advice by interpreting the received data. Any loss in this data can put patient's life in danger. Existing mobility schemes suffer from packet loss during HO in HWSN and results in loss of data. The packet loss is due to the termination of connection during HO. Attempts should be made to control the packet loss during HO by buffering the packets either at previous AP (The current home of MN) or at new AP (where the MN intends to move) and deliver the buffered packets to the MN after the HO is completed. In this way the critical data can be saved from being lost and QoS improved in HWSN mobility scenario.

## 4. Proposed Area and Solution:

From the survey of literature available on mobility management schemes we conclude that the existing schemes still suffer from extra signaling cost and packet loss during Handoff which is not acceptable when some critical data is being transferred. In our case study of HWSN, patient's mobility can be tracked by deploying sensor devices on the body of patient, thus creating wireless body area network (WBAN) [71]. These sensor nodes (Bio Sensors) are attached to patient's body to get the health related data of the patient. Because of the time criticality of this data, the patient needs immediate expert's (Doctor) advice. No loss of data packets is acceptable in such healthcare scenarios [38]. Earlier attempts [22, 19, 41] to solve this problem are PMIPv6 based sensor's mobility, still suffer from packet loss and extra signaling cost because there exists a period when the MN is unable to send or receive packets during HO. Also PMIPv6 protocol operations suffering from handover latency and data loss due to the connection termination during HO. Thus, to reduce the handover latency and data loss in PMIPv6 based applications where data is critical, an approach to buffer the data packets before the Handoff (HO) takes place should be made and after completion of HO these buffered packets can be delivered without loss of information and the patient can be monitored efficiently by an expert sitting either at Hospital Monitoring station (HMS) in the hospital premises or at a far place. The best suitable protocol to monitor the patient's mobility (patient moves in different departments inside the hospital under different MAG domains) in hospital scenario would be Fast Handovers for PMIPv6 (FHPMIPv6) [72]. FHPMIPv6 buffers the data packets either at Previous Mobile Access Gateway (pMAG) or New MAG (nMAG) and these packets are delivered to the Mobile Node (MN) after the HO Process is over. Fast Handovers for PMIPv6 (FHPMIPv6) has been standardized by the IETF in RFC 5949 [72] and lot of work is being carried out using this protocol to avoid packet loss during HO process. In [34] FHPMIPv6 used for highway vehicular traffic scenario. In [25] same protocol used to provide seamless handover in high speed trains and produced better results than PMIPv6. To the best of our knowledge FHPMIPv6 has not been used for monitoring patient's mobility in HWSN so far.

**4.1. FHPMIPv6 overview:** The protocol is aimed at reducing the HO latency and Packet loss. The idea behind the protocol is based on:

1. A handover can be predicted and the new MAG can be triggered to send a proxy binding update to the LMA, so that downlink traffic from LMA to MN can already be switched to the new MAG. The new MAG will buffer the traffic until the MN is connected after HO.
2. Downlink traffic arriving at the old MAG will be buffered if the MN has disconnected and redirected to the new MAG. The new MAG will inform the old MAG of the handover by means of a Fast Binding Update (FBU) message to instruct the old MAG to redirect traffic.

For our case study of HWSN, where the sensors are deployed on the patient's body form a group. The Signaling flow diagram for patient's mobility in HWSN is given in figure 10.

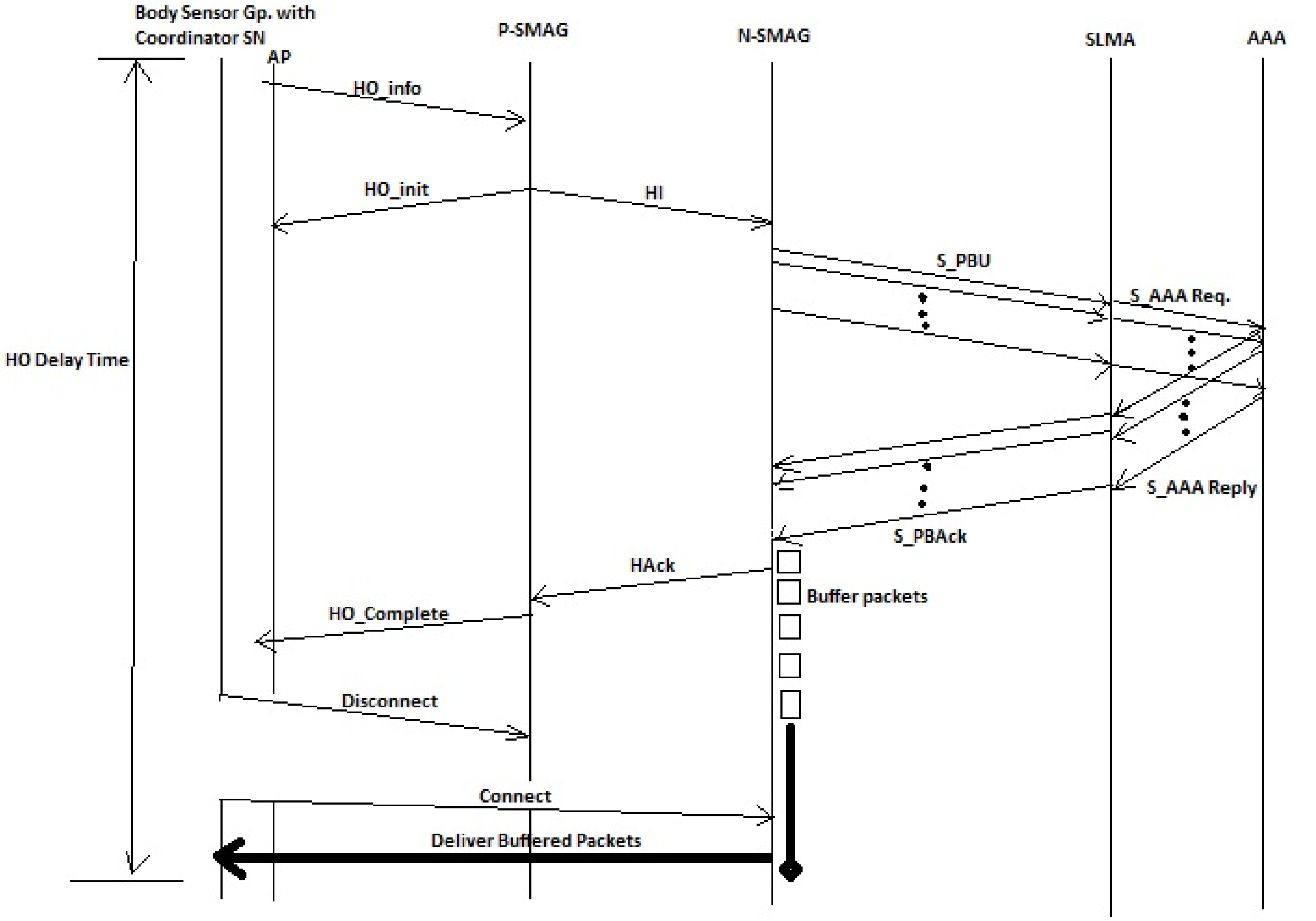

Fig. 10: Basic Signaling Flow Diagram of FHPMIPv6

**4.1.1. The Registration Phase:** The registration process aims at reducing the number of control message transfer (see fig. 10). One of the body sensors from group behave as a coordinating Node which interacts with rest of the body sensors in 6LoWPAN environment. The procedure of registration given as below:

1. The P_SMAG receives a periodic notification from L2 (from the current Access Point or indirectly the MN) that a handover is imminent (To check HO is imminent, see algo for connection quality comparison below). The L2 notification includes the information about expected new SMAG (N_SMAG) to which Coordinating SN will connect (HO_info). The address of the MAG can be broadcasted by the connected APs on their pilot channel as the MAGs are provide with table relating the APs of its neighboring MAGS, so the coordinating SN can inform the P_SMAG (HO_info).

2. The P_SMAG can easily derive the address of N_SMAG (Procedure for deriving the address is outside the scope of paper).
3. P_SMAG decides to initiate the handover and informs its AP (L2 message HO_init). Also it sends handover initiation (HI) message (MN's identifier and timestamp) to the N_SMAG.
4. Upon receiving HI message, N_SMAG sends a Sensors proxy binding update (S_PBU) message to SLMA by using the timestamp in HI message. SLMA sends S_AAA request to AAA server for authentication and gets S_AAA reply as acknowledgement.
5. The SLMA installs the new binding information in its cache and returns a Sensors proxy binding acknowledgement (S_PBAck) message to the N_SMAG. Then SLMA forwards all traffic intended for Coordinating MSN (mobile sensor node) to N_SMAG.
6. Upon receiving the S_PBAck message, the N_SMAg sends handover acknowledgement (HAck) message to the P_SMAG and initiate buffering of packets for the Coordinating MSN until the MSN connects to the N_SMAG through new AP after which N_SMAG forwards all buffered traffic to MSN.
7. The P_SMAG after receiving HAck message, sends a L2 HO_Complete message to its AP to instruct the MSN to get disconnected from old AP and connect to the new one, if it has not done already.
8. After joining the new AP of N_SMAG, MSN sends NDP (neighbor discovery protocol) request message to the N_SMAG to use the address auto configuration service of IPv6 and starts receiving the traffic from SLMA through N_SMAG.

**4.2. Connection Quality Comparison Algorithm:** As FHMIPv6 is based on advance prediction of HO. To know whether to perform HO or not depends upon the connection quality of the signal generated by the APs of SMAGs. The algorithm for connection quality comparison is:

1. Mobile Sensor Node (MSN) receives L2 (RA) notification message from new SMAG (N_SMAG) through periodic Router Advertisement messages.
2. First legitimacy of advertisement message is checked whether the advertising SMAG is registered with SLMA or not (by checking the home network prefix (HNP)).
3. If not registered, do nothing and go to step 7 else go to step 4.
4. Check signal quality of new AP of N_SMAG and see whether it is better than the signal quality of P_SMAG's AP. Calculate if Signal strength (N_SMAG) – Signal strength (P_SMAG) > X, where X is some predefined threshold value for signal quality of AP used to perform HO. Exact value of X is outside the scope of this paper but it should be kept as minimum as possible to avoid Packet Loss.
5. If step 4 is true, MSN disconnected from P_SMAG's AP and connected to N_SMAG's AP and a message request to delete the P_SMAG registration with SLMA is sent to SLMA to avoid further queries about P_SMAG by SLMA and go to step 7. Otherwise go to step 6.
6. MSN remains connected to the current AP of P_SMAG.
7. Look for another neighbor AP or SMAG advertisement, if found go to step 1 else step 8.
8. Continue traffic forwarding through the current SMAG.

**4.3. Performance Analysis:** for the analysis of packet loss and HO latency in our case study of HWSN, certain notations have to be used. Packet loss is defined as number of packets lost during HO period of patient and is analyzed here as the time interval during which generated packets are lost. HO latency is the period from the moment when coordinating sensor node is not able to receive the packets from P_SMAG to the moment when it receives the first packet from N_SMAG (i.e. the connection termination time because of HO). For the analysis of HO latency and Packet loss, the notations used are given in the table 2.

TABLE 2: Notations used for Analysis

| | |
|---|---|
| n | Number of Body Sensors used on patient' body |
| $HO_{PL}$ | Packet loss during handover due to untimely HO prediction |
| $HO_{Lat}$ | Handoff latency caused by signaling messages delay intervals |
| $D_{SMAG-AP}$ | Delay time between SMAG and its AP |
| $D_{MAG-MAG}$ | Delay time MSN takes between two MAGS during movement (HO) |
| $T_{U_Pred}$ | Time taken for HO untimely prediction |
| $D_{S_PBU}$ | Delay for sending the n Sensor's binding update to SLMA |
| $D_{S_PBAck}$ | Delay for acknowledging S_PBU by SLMA |
| $D_{S_AAAreq}$ | Delay for sending n authentication request to AAA Server |
| $D_{S_AAAreply}$ | Delay to acknowledge S_AAAreq request |
| $D_{L2}$ | Time taken to advertise the neighboring AP's information |
| $D_{DHCP}$ | Time elapsed for Address configuration |

Let us first analyze the packet loss during HO. If the handover prediction happens timely then there will be no packet loss, since all arriving packets at P_SMAG will be delivered to Mobile Sensor Node's group and the packets arriving at N_SMAG will be buffered until MSN gets connected to the N_SMAG. If the HO prediction does not happen timely, packets that are already sent by the P_SMAG to its AP before the P_SMAG gets notified of disconnection of MSN get lost. This packet loss is mainly during delay time between P_SMAG and its AP. If in this case the HO prediction had done little late then the packet loss could be little lesser as some more packets would have delivered to MSN before its disconnection. The handover packet loss because of untimely HO prediction is dependent on different delay factors, shown in the equation below:

$$HO_{PL} = \sum^{n}_{k=1} k(D_{SMAG\text{-}AP}) + D_{MAG\text{-}MAG} + T_{U_Pred} \quad (1)$$

$$HO_{PL} = n(D_{SMAG_AP}) + D_{MAG\text{-}MAG} + T_{U_Pred} \quad (2)$$

$$\text{Avg. } HO_{PL} = HO_{PL} / n = D_{SMAG_AP} + (D_{MAG\text{-}MAG} + T_{U_Pred}) / n \quad (3)$$

The handover latency is contributed due to some delay intervals and signaling messages used during handover.

$$HO_{Lat} = \sum^{n}_{k=1} i(D_{S_PBU} + D_{S_PBAck} + D_{S_AAAreq} + D_{S_AAAreply}) + D_{L2} \quad (4)$$

$$HO_{Lat} = n(D_{S_PBU} + D_{S_PBAck} + D_{S_AAAreq} + D_{S_AAAreply}) + D_{L2} \quad (5)$$

$$\text{Avg. } HO_{Lat} = HO_{Lat} / n = D_{S_PBU} + D_{S_PBAck} + D_{S_AAAreq} + D_{S_AAAreply} + D_{L2} / n \quad (6)$$

By assuming AAA server to be included in SLMA, we can rewrite the eq. 6 as

$$\text{Avg. } HO_{Lat} = D_{S_PBU} + D_{S_PBAck} + D_{L2} / n \quad (7)$$

In the analysis shown above, we did not included $D_{DHCP}$ (time elapsed for address configuration of Sensor nodes) delay because of the address auto configuration [73] property of IPv6. Also by comparing our analysis with the analysis given in [41], we can say that our proposal reduces signaling cost and packet loss. Therefore FHPMIPv6 for 6LoWPAN WSN mobility is best suitable particularly in the areas where critical data has to be protected.

## 5. Conclusion

In this article, we presented a brief description of host based mobility management protocols along with a complete study of network based mobility and 6LoWPAN WSN mobility management schemes. Further we presented some of the challenging areas in mobility management where improvements are still needed. One such area highlighted in this paper is healthcare (HWSN), where patient's mobility is continuously monitored inside the hospital premises. During patient's mobility (HO), time critical data (health parameters) is always being transmitted to the monitoring station. That data has to be protected from being lost during HO. Already existing mobility approaches [63, 74, 75] are tunnel based where sensor nodes have to send lot of signaling messages, therefore not suitable for 6LoWPAN mobility. [41] Used PMIPv6 based mobility approach for sensor's group mobility which provides better results in comparison to original protocol but that still suffers from packet loss and there is further scope for reducing signaling cost. Therefore, the suggested FHPMIPv6 is best suitable for reducing packet loss. The analysis of FHPMIPv6 for HWSN scenario shows reduction in packet loss and signaling cost. Also in this paper we presented Connection Quality Comparison algorithm used for predicting HO. The algorithm is useful in avoiding unnecessary handovers by comparing connection quality to some predefined threshold value. In future, an attempt would be made for practical implementation of FHPMIPv6 for some critical areas in WSN. Also a study of security issues in 6LoWPAN WSN will be performed with emphasis on Biometric security in 6LoWPAN WSN mobility as very less work has been done on Biometric security in 6LoWPAN.